# Bekestein Bound of Information Number *N* and its Relation to Cosmological Parameters in a Universe with and without Cosmological Constant

Ioannis Haranas[1] Ioannis Gkigkitzis[2]

[1] Department of Physics and Astronomy, York University
4700 Keele Street, Toronto, Ontario, M3J 1P3, Canada
E-mail:yiannis.haranas@gmail.com

[1] Departments of Mathematics, East Carolina University
124 Austin Building, East Fifth Street, Greenville
NC 27858-4353, USA
E-mail: gkigkitzisi@ecu.edu

**Abstract**
Bekenstein has obtained is an upper limit on the entropy *S*, and from that, an information number bound *N* is deduced. In other words, this is the information contained within a given finite region of space that includes a finite amount of energy. Similarly, this can be thought as the maximum amount of information required to perfectly describe a given physical system down to its quantum level. If the energy and the region of space are finite then the number of information *N* required in describing the physical system is also finite. In this short letter two information number bounds are derived and compared for two types of universe. First, a universe without a cosmological constant lamda and second a universe with a cosmological constant lamda $\Lambda$ are investigated. This is achieved with the derivation of two different relations that connect the Hubble constant and cosmological constants to the number of information *N*. We find that the number of information *N* involved in a the two universes are identical or $N_2 = N_{2_\Lambda}$, and that the total mass of the universe scales as the square root of the information number *N*, containing. an information number *N* of the order of $10^{122}$. Finally, we expressed Calogero's quantization action as a function of the number of information *N*. We also have found that in self-gravitating systems the number of information *N* in nats is the ratio of the total kinetic to total thermal energy of the system.

## 1. Introduction

There is an upper bound for the ratio of the entropy *S*, to the energy *E* of a mass *M*, for any bounded system with an effective radius *R*. This is known as the Bekenstein (1981) upper bound given by the relation:

$$S \leq 2\pi \left( \frac{k_B R E}{\hbar c} \right). \tag{1}$$

Where $\hbar$ is Planck's constant, and c the speed of light. Using Eq.(1) and the equation given in Haranas and Gkigkitzis (2013), for the entropy *S* to be (Lloyd, 2001):

$$S = k_B N \ln 2, \tag{2}$$

___________________________________

In this paper both authors have contributed equally.

where $k_B$ is the Boltzmann constant Eqs. (1) and (2) we obtain that the number of information $N$ in nats contained in the quantum states in the sphere is given by the equation:

$$N \leq \frac{2\pi}{\ln 2}\left(\frac{RE}{\hbar c}\right). \tag{3}$$

Where the number of information $N$ is given in nats. Taking, $R = c/H_0$ to be the Hubble radius and assuming that the universe self gravitational energy $E_{gr} = \hbar H_0$ (Johri, 1996), we find that the number of information in nats $N$ is equal to:

$$N \leq \frac{2\pi}{\ln 2}\left(\frac{E}{\hbar H_0}\right) = \frac{2\pi}{\ln 2}\left(\frac{E}{E_{gr}}\right) = \frac{2\pi}{\ln 2}\left(\frac{\omega}{H_0}\right) = \frac{2\pi}{\ln 2}\left(\frac{R_{gr_u}}{2\lambda_{c_u}}\right), \tag{4}$$

Similarly Eq. (4) can be written as the ratio of the angular frequency of oscillation of the universe over its Hubble constant $H$, or the ratio of gravitational radius of the universe over the universe's Compton wavelength. Next we find the entropy at the Hubble horizon to be:

$$S_H = \frac{k_B c^3}{4G\hbar} A_H = \frac{\pi k_B c^5}{G\hbar H_0^2} = \pi k_B \left(\frac{c}{H_0 \ell_P}\right)^2. \tag{5}$$

With reference to Haranas and Gkigkitzis (2013) and equating Eq. (2) to (5) we obtain Hubble's parameter as a function of the number of information $N$ to be:

$$H_0 = \sqrt{\frac{\pi}{\ln 2}}\left(\frac{c}{\ell_P \sqrt{N}}\right), \tag{6}$$

Substituting Eq. (6) in (4) and given that $N$ is positive we obtain that:

$$N^2 \leq \frac{4\pi}{\ln 2}\left(\frac{E^2 \ell_P^2}{\hbar^2 c^2}\right) N. \tag{7}$$

Solving Eq. (7) for $N$ we obtain that:

$$0 \leq N \leq \frac{4\pi}{\ln 2}\left(\frac{E^2 \ell_P^2}{\hbar^2 c^2}\right) = \frac{4\pi}{\ln 2}\left(\frac{E}{E_P}\right)^2 = \frac{4\pi}{\ln 2}\left(\frac{M_U}{m_P}\right)^2, \tag{8}$$

where, and $E_P = \sqrt{\frac{\hbar c^5}{G}}$ is the Planck energy, $m_P = \sqrt{\frac{\hbar c}{G}}$ is the Planck mass, and $\ell_P = \sqrt{\frac{\hbar G}{c^3}}$ is the Planck length. Therefore we find that the number of information $N$ at the horizon of the Friedman universe is just the square of ratio of the total energy of the universe that is attributed to the universe's total mass to that of the Planck energy, or equivalently to the square of the total mass of the universe to that of the Planck mass. Similarly, in a universe with zero curvature and cosmological constant $\Lambda$ and with reference to Haranas and Gkigkitzis (2013) we have that the cosmological constant can be written as a function of the information number bit $N$ in the following way:

$$\Lambda = \left( \frac{3\pi}{N \ln 2 \ell_P^2} \right) = \frac{3\pi \Lambda_{max}}{\ln 2 N}, \qquad (9)$$

where $\Lambda_{max} = 1/\ell_P^2 = c^3/G\hbar$ (Haranas, 2002) the maximum cosmological constant, $N$ the number of information in nats, $\ell_P$ is the Planck length. Therefore writing the cosmological de-Sitter horizon as function the information $N$ we obtain:

$$r_{H_{1,2}} = \sqrt{\frac{3}{\Lambda}} = \ell_P \sqrt{\frac{N \ln 2}{\pi}}. \qquad (10)$$

Substituting Eq. (10) into first part of Eq. (4) after simplifying in a way similar to Eq. (6) we obtain that:

$$N^2 \leq \frac{4\pi}{\ln 2} \left( \frac{E^2 \ell_P^2}{\hbar^2 c^2} \right) N, \qquad (11)$$

from which we obtain that:

$$0 \leq N_\Lambda \leq \frac{4\pi}{\ln 2} \left( \frac{E^2 \ell_P^2}{\hbar^2 c^2} \right) = \frac{4\pi}{\ln 2} \left( \frac{E}{E_P} \right)^2 = \frac{4\pi}{\ln 2} \left( \frac{M_U}{m_P} \right)^2, \qquad (12)$$

and therefore we obtain that:

$$N_2 = N_{2_\Lambda}, \qquad (13)$$

in other words the number of information contained in a Friedmann universe without cosmological constant results to a Bekenstein information bit bound that is identical to the universe involving a cosmological constant $\Lambda$, or symbolically $N = N_{2_F} = N_{2_\Lambda}$. This is the number of information in nats with the help of which information can be decompressed through matter and energy, for the two types of universe. We find that for both types of universe the number of information $N$ depends on the same fundamental parameters i.e. the energy due to mass $M$ over the Planck energy $E_P$ or equivalently the mass of the universe $M$ over the Planck mass $m_P$. From Eq. (4) we see that the number of information $N$ constitutes the "connecting entity" of various fundamental cosmological parameters. As a result from Eq. (4) we obtain the gravitational radius of the universe can be expressed in the following way:

$$R_{gr_U} = \left( \frac{2 E \lambda_{c_U}}{\hbar H_0} \right). \qquad (14)$$

Next, substituting for the universe's Compton wavelength $\lambda_{c_U} = \hbar / M_U c$, and $R_{gr_U} = 2GM/c^2$ and $E = M_U c^2$ we obtain that the mass of the universe is given by:

$$M_U = \frac{c^3}{GH_0}. \tag{15}$$

This is an equation similar to that predicted by Hoyle (1958), Narlikar (1993), Carvalho (1995), Haranas and Harney (2009) as well as (Valev, 2010). For example Valev (2010) obtains the same equation via a dimensional analysis. Having expressed $H_0$ as a function of the number of information $N$ in nats, we can use Eq. (6) to express the mass of the universe as a function of the number of information $N$ in the universe without cosmological constant to be:

$$M_U = \frac{c^2 \ell_P}{G}\sqrt{\frac{\ln 2N}{\pi}} = 0.470 \frac{c^2 \ell_P}{G}\sqrt{N} = 0.470 \frac{\Lambda_{max} \hbar \ell_P}{c}\sqrt{N} = 0.470 \frac{\hbar}{c\ell_P}\sqrt{N}. \tag{16}$$

Finally, using that the Planck momentum can be written as $m_P c = \hbar/\ell_P$ from the last expression of Eq. (18) we can obtain that:

$$M_U = 0.470\, m_P \sqrt{N}. \tag{17}$$

Similarly, in the universe with cosmological constant $\Lambda$ we obtain that:

$$M_U = \frac{c^2\sqrt{3}}{G\sqrt{\Lambda}} = \sqrt{\frac{\ln 2}{\pi}}\frac{c^2 \ell_P}{G}\sqrt{N} = 0.470 \frac{\Lambda_{max} \hbar \ell_P}{c}\sqrt{N} = 0.470 \frac{\hbar}{c\ell_P}\sqrt{N} = 0.470\, m_P \sqrt{N} \tag{18}$$

The first tem in the RHS of Eq. (18) agrees with the Eq. (8) given in Capozziello and Funkhouser (2009) and Eq. (3.8) in Funkhouser (2008). We express this equation as a function of information $N$, from which the number of information $N$ in nats related to the universe's total mass can be calculated. Therefore taking $M_U \cong 1\times 10^{53}$ kg, (Immerman, 2001) and $m_P = 2.176 \times 10^{-8}$ kg obtain that:

$$0 \leq N \leq 4.526\left(\frac{M_U}{m_P}\right)^2 \leq 0.996 \times 10^{122} \text{ nats}. \tag{19}$$

The number $10^{122}$ appears in an ensemble of pure numbers naturally produced from fundamental cosmological parameters that might constitute a new-large number coincidence similar to that postulated by Dirac (1973). These numbers constitute a compelling, new large number coincidence problem (Funkhouser, 2008) and (Haranas and Gkigkitzis, 2013). This is possible after the derivation of two relations which connect the cosmological constant $\Lambda$ and the Hubble constant $H_0$ to the information number $N$. We find that the total mass of the universe has a $\sqrt{N}$ dependence on the information number bit $N$. Furthermore, we note that the mass of the universe expressed in terms of fundamental parameters that basically become coefficients of $\sqrt{N}$ term and carry units of mass. We see that for $N=1$ the universe achieves a mass $M_U = 0.470\, m_P$, for which Eq. (22) gives $N = 0.999 \approx 1.0$. Inflationary cosmological scenarios tell us that at time $t = t_P$ then $m = m_P$, (Linde, 1990) and therefore we can say that a universe with $M_U = 0.470\, m_P$ will imply that the first nat of information is decompressed through matter and energy at a time $t$ earlier that the Planck time. In Capozziello et al., (2001) the authors examine self-gravitating systems, where they consider virialized systems, giving the equation for the total energy of the system to be:

$$2E_{kin} + U = 0, \tag{20}$$

where $E_{kin}$ is the kinetic energy, and $U$ the gravitational energy. Therefore, the total energy $E$ given by Capozziello, (2001) is given by:

$$E \cong E_{kin} = N_0 mv^2. \tag{21}$$

Using Capozziello's equation for the total energy, we write the entropy of the self-gravitating system to be:

$$S \cong \frac{N_0 mv^2}{T}, \tag{22}$$

and using Eq. (2) and (22) we obtain the number of information in nats related to the self-gravitating system in the following way:

$$N = \frac{N_0 mv^2}{T k_B \ln 2} = 2 \frac{N_0}{\ln 2} \left( \frac{E_{kin}}{E_{th}} \right) = \frac{2 E_{kin}}{T \Delta S}, \tag{23}$$

where $N_0$ is the total numbers of bodies of mass $m$ contained within the self-gravitating system, and $E_{th}$ is the total thermal energy of the system, and $\Delta S = k_B \ln 2$ is the change in entropy. Similarly, using Eq. (23) we obtain an expression for the characteristic (minimal) unit of action $\alpha = \varepsilon \tau$ (Capozziello, et al., 2001) per granular component to be:

$$\alpha = \varepsilon \tau \cong \frac{A}{\left[ \frac{k_B T}{mv^2} N \ln 2 \right]^{3/2}} = \frac{A}{\left[ \left( \frac{E_{kin}}{E_{th}} \right) N \ln 2 \right]^{3/2}}. \tag{24}$$

Finally, with reference to Calogero's work on cosmic quantization (Calogero, 1997) the author predicts a quantum of action $h$ as a function of basic parameters of physics, namely:

$$h \cong G^{1/2} m^{3/2} R^{1/2}, \tag{25}$$

where $G$ is the gravitational constant, $m$ mass of the nucleon or hydrogen atom, and $R$ is the radius of the universe (Calogero, 1997). We can show the relation of Calogero's quantum of action $h$, to number of information $N$ in nats by taking the radius of the universe at the Hubble horizon in two kinds of different universes. First, at Hubble horizon of a Friedmann universe we have that $R = c / H_0$ and therefore we obtain that:

$$h \cong G^{1/2} m^{3/2} \left( \frac{c}{H_0} \right)^{1/2} \tag{26}$$

Upon substituting Eq. (6) into Eq. (26) and eliminating the Planck length we obtain that:

$$h \cong \left( \frac{\ln 2}{\pi} \right)^{1/4} (G m^3 \ell_p)^{1/2} N^{1/4} = \left( \frac{\ln 2}{\pi} \right)^{1/4} (G m^3)^{1/2} \left( \frac{Gh}{2\pi c^3} \right)^{1/4} N^{1/4}, \tag{27}$$

solving for $h$ we obtain the following real solutions

$$h_1 = h \cong 0, \tag{28}$$

$$h_2 = h \cong 0.466 \left( \frac{\ln 2}{2} \right)^{1/3} \left( \frac{Gm^2}{c} \right) N^{1/3}. \tag{29}$$

Finally, in a de-Sitter universe with cosmological constant we have that Calogero's equation becomes:

$$h \cong G^{1/2} m^{3/2} \left( \sqrt{\frac{3}{\Lambda}} \right)^{1/2} = G^{1/2} m^{3/2} \left[ \ell_P \left( \frac{N \ln 2}{\pi} \right)^{1/2} \right]^{1/2}, \tag{30}$$

thus eliminating the Planck length $\ell_P$ and solving for $h$ we obtain the following real values for the quantum of action:

$$h_1 = h \cong 0, \tag{31}$$

$$h \cong 0.466 \left( \frac{\ln 2}{2} \right)^{1/3} \left( \frac{Gm^2}{c} \right) N^{1/3}. \tag{32}$$

Using Eq. (29) and solving for $N$ we obtain that:

$$N = \frac{2\pi^2}{\ln 2} \left( \frac{c^3}{G^3 m^6} \right) h^3 = \frac{16\pi^5}{\ln 2} \left( \frac{c\hbar}{Gm^2} \right)^3 = \frac{16\pi^5}{\ln 2} \left( \frac{m_P}{m} \right)^6, \tag{33}$$

Therefore, the number of information $N$ in Calogero's scheme of cosmic quantization depends upon the ratio of two fundamental masses i.e. the Planck mass and mass of the nucleon raised to the sixth power. Similarly, we can say that Calogero's quantum of action $h$ as related to the universe's radius calculated at the Hubble radius, involves an $N^{1/3}$ dependence on the number of information in nats. Using $m_P = 2.176 \times 10^{-8}$ kg and $m = 1.670 \times 10^{-27}$ kg (Calogero, 1997) we obtain that the number of information in nats is:

$$N = 3.457 \times 10^{118}, \tag{34}$$

$10^{118}$ is a large number that appears in an ensemble of pure numbers that they are naturally produced from fundamental cosmological parameters, that they are probably part of a new-large number hypothesis, similar to the one postulated by Dirac. In the case where $m = m_P$ we obtain that:

$$N = \frac{16\pi^5}{\ln 2} \cong 7064 \ . \tag{35}$$

Similarly, in the case of a self-gravitating system, let us consider the typical open star cluster Pleiades for which $M = 300 M_{solar}$ $R = 3.5$ pc, $N_0 = 300$ stars and $\sqrt{\bar{v}^2} \cong 4.63 \times 10^{-2} \sqrt{N_0 m/R}$ km/s (Chandrasekhar, 1960), we obtain that $\bar{v} = 0.43$ km/s, with temperatures in the range $4000 \text{ K} \leq T \leq 15000$ K. Using an average temperature $\bar{T} = 10000$ K, we obtain that the number of information $N$, is:

$$N = 1.154 \times 10^{57} \ . \tag{36}$$

## 2. Conclusions

In this letter we have used Bekenstein's upper bound of entropy, to calculate an upper bound for the number of information bits $N$ in two different types of universe. We find that the number of information $N$ in a universe without a cosmological constant is identical to the number of information $N$ in a universe with a cosmological constant lamda. Furthermore, we find that the mass of the universe can be expressed in terms of many various cosmological parameters that basically become the coefficients of the $\sqrt{N}$ term and have units of mass, thus the mass of the universe $M_U \propto m_P \sqrt{N}$ and that the universe achieves a mass $M_U = 0.470 \, m_P$ if $N = 1$. Finally, we have expressed Calogero's quantization action as a function of the number of information $N$. We have found that the action $h$ as related to the radius of the universe, when taken at the universe's horizon, involves a large number of information in nats, that is the same in a universe with or without a cosmological constant. In relation to Capozziello's results for self-gravitating systems, we have found that the number of information $N$ in nats is the ratio of the total kinetic to total thermal energy of the system.